\pdfoutput=1
\documentclass[aps,prd,a4paper,11pt,superscriptaddress,nofootinbib]{revtex4-2}
\usepackage{microtype}
\usepackage{graphicx}
\graphicspath{{fig/}}

\usepackage{mathtools}
\usepackage{amssymb}
\DeclareMathOperator{\sgn}{sgn}
\DeclarePairedDelimiter{\abs}{\lvert}{\rvert}

\usepackage[dvipsnames]{xcolor}
\usepackage[T1]{fontenc}
\usepackage[pdfusetitle,bookmarksopen,colorlinks,allcolors=blue]{hyperref}

\begin{document}
\title{Scattering of compact kinks}

\author{F.~M.~Hahne}
\email{fernandomhahne@gmail.com}

\author{P.~Klimas}
\email{pawel.klimas@ufsc.br}

\affiliation{Departamento de F\'isica, Universidade Federal de Santa Catarina,\\
Campus Trindade, 88040-900, Florian\'opolis, Brazil}

\begin{abstract}
We study the scattering processes of kink-antikink and kink-kink pairs in a field theory model with non-differentiable potential at its minima.
The kink-antikink scattering includes cases of capture and escape of the soliton pair separated by a critical velocity, without windows of multi bounce followed by escape.
Around the critical velocity, the behavior is fractal.
The emission of radiation strongly influences the small velocity cases, with the most radiative cases being also the most chaotic.
The radiation appears through the emission of compact oscillons and the formation of compact shockwaves.
The kink-kink scattering happens elastically, with no emission of radiation.
Some features of both the kink-antikink and the kink-kink scattering are explained using a collective coordinate model, even though the kink-kink case exhibits a null-vector problem.
\end{abstract}

\maketitle

\section{Introduction}

Topological solitons in scalar field theories are of interest in a vast range of physical phenomena, ranging from effective models of fundamental interactions, passing through condensed matter systems, to cosmological applications.
Although having localized energy density, solitons in most field theory models have infinite size, reaching the vacua only in an asymptotic manner, i.e.\ they exhibit exponential tails.
However, there are some examples of solitons contained to a compact support, called compactons.

The term compacton has been first introduced in the context of a Korteweg–De Vries equation~\cite{Rosenau:1993zz}.
Since then, compactons have been discovered in models with non-analytic potential, first discussed in the study of interacting pendulums with solid barriers limiting its motion~\cite{Arodz:2002yt,Arodz:2003mx}.
A potential is said non-analytic when its one-sided derivatives around the minima are non-zero and with different values from the left than from the right.
Potentials of this kind approach its minima with a characteristic V-shape.
The sharpness of the potential has as consequence the existence solutions with compact support.
In particular, instead of having exponential tails, compactons in field theory models with non-analytic potential approach the vacuum parabolically.

Several fascinating features of models with V-shaped potentials have been previously discussed.
For example, the radiation spectrum of such models has been shown to be dominated by compact oscillons, which can scatter and generate new oscillons in a pattern that is likely fractal~\cite{Hahne:2019ela}.
We have also previously discovered that compact topological kinks allow the propagation of signals through its bulk with the speed of light, and that they can scatter with compact oscillons, i.e.\ radiation, without exchanging momentum~\cite{Hahne:2022wyl}.

The interaction of topological defects has been an active area of research for decades, and one where several important phenomena are still not fully understood.
For a review of the subject, see ref.~\cite{Kevrekidis:2019zon}.
Except in the rather special case of integrable models, a full field theory description of the interaction between kinks is complicated, among other things, by the presence of radiation.
Nonetheless, there has been progress in the description of scattering of kinks in theories such as the $\phi^4$ Klein-Gordon model by the use of a collective coordinate approximation~\cite{Manton:2021ipk,Manton:2020onl,Pereira:2020jmi,Pereira:2021gvf,Adam:2021gat,Adam:2023qgx,Blaschke:2023mxj}.

In this work, we focus on the scattering of compact kinks in a model with non-analytic potential.
First, we present the model and compacton solution in section~\ref{sec:compact-kinks}, where we also use the collective coordinate approximation to describe the dynamics of a single kink.
In section~\ref{sec:kink-antikink-scattering} we study the kink-antikink scattering, first through the analysis of numerical results and then by the collective coordinate approximation, using a positional and an internal mode for each compacton.
We apply the same approach to the kink-kink scattering in section~\ref{sec:kink-kink-scattering}, were we discuss difficulties arising from a null-vector problem in the collective coordinate description.
We finish by summarizing our results in section~\ref{sec:conclusions}.

\section{Compact kinks}
\label{sec:compact-kinks}

\begin{figure}
    \centering
    \includegraphics{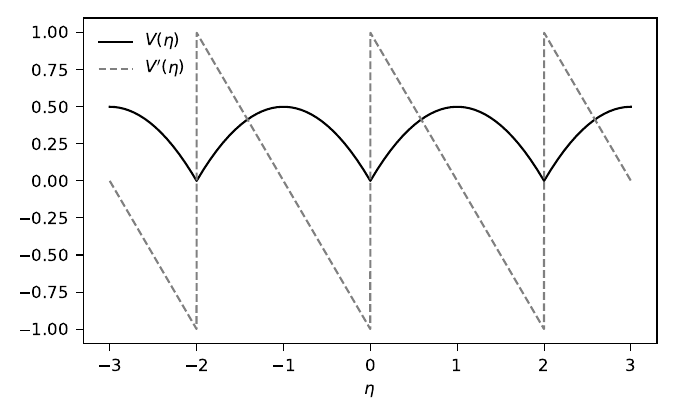}
    \caption{The potential $V(\eta)$ and its derivative $V'(\eta)$.}
    \label{fig:potential}
\end{figure}

Let us start by introducing the relevant model and topological kink solutions.
In this work we consider the theory of a real scalar field $\eta$ in $1+1$ dimensional model described by the action
\begin{equation}
    S = \int dt \, dx \left[\frac{1}{2}(\partial_t\eta)^2-\frac{1}{2}(\partial_x\eta)^2-V(\eta)\right] \label{eq:action}
\end{equation}
where the potential $V(\eta)$ is non-analytic at its minima, i.e.\ its left and right one-sided derivatives have different values around the vacuum.
In units natural to our model, the potential minima are located at even integer values of the field and are connected by parabolas, as in figure~\ref{fig:potential}.
One possible way of expressing this potential is the following
\begin{align*}
    V(\eta) &= \sum_{n=-\infty}^{\infty} \left( |\eta - 4n| - \frac{1}{2}(\eta - 4n)^2 \right) H_n(\eta), \\
    H_n(\eta) &\coloneqq \theta(\eta - 4n + 2) - \theta(\eta - 4n - 2)
\end{align*}
where $\theta$ is the Heaviside step function.
The function $H_n(\eta)=1$ for $|\eta-4n|<2$ and $H_n(\eta)=0$ otherwise.
In the context of this formula, the minima are $\eta = 4n$ and $\eta = 4n \pm 2$, where $n \in \mathbb{Z}$ labels the segment of the potential.
A simpler expression for the minima is $\eta = 2k$, where $k \in \mathbb{Z}$ labels the minima.
The Euler--Lagrange equation is the modified Klein--Gordon equation
\begin{equation*}
    \partial_t^2 \eta - \partial_x^2 \eta + V'(\eta) = 0
\end{equation*}
where $V'(\eta)$ is the derivative of the potential.
Note that the classic derivative of $V(\eta)$ is ill-defined for $\eta = 2k$, however these values correspond to static fields configurations at the minima of the potential with zero energy, and are physically acceptable.
Therefore, we must extend our definition of $V'(\eta)$ so that $V'(2k) \coloneqq 0$.
This definition agrees with the weak derivative of $V(\eta)$.
Explicitly, we write
\begin{equation*}
    V'(\eta) = \sum_{n=-\infty}^\infty \left[\sgn(\eta - 4n) - (\eta - 4n) \right] H_n(\eta)
\end{equation*}
with the requirement that $\sgn(0)=0$.

It is important to emphasize that this model is related to real physical systems.
For instance, in mechanical system of coupled pendulums where the motion is limited by rigid barriers, the barriers limit the possible angles of the pendulums.
In the continuum limit this corresponds to a limitation in the values of the field~\cite{Arodz:2002yt}.
The limitation of the values of the angle can be dealt with by the use of the so-called ``unfolding'' transformation, yielding a model with potential made up of pieces of the cosine function~\cite{Arodz:2005}.
If we further impose that the barriers limit the pendulums to small angles, the cosine function can be approximated by a quadratic function.
In this case we get a displaced and rescaled version of our model.

Our model is also related to the Skyrme model, since the Skyrme model can be written as the combination of two Bogomol'nyi-Prasad-Sommerfeld (BPS) submodels~\cite{Adam:2017pdh}.
The angular part of the first of this submodels can be solved through a rational map ansatz, while the radial part is equivalent to a field theory in $1+1$ dimensions, with a limitation to the possible field values.
This limitation can be avoided by an ``unfolding'' transformation.
This results in our model, as discussed in the ref.~\cite{Klimas:2018woi}.

In the static case, the action is minimized by solutions of the BPS equations~\cite{Bogomolny:1975de,Prasad:1975kr}
\begin{equation*}
    \frac{d\eta}{dx} = \pm \sqrt{2V(\eta)}.
\end{equation*}
When this equation is integrated with non-trivial boundary conditions, this equations give rise to kink and antikink solutions interpolating the different vacua.
In particular for our model, such soliton solutions are different from the minima of the potential only inside a compact support, i.e.\ they are compactons.
This is a consequence of the non-analytic character of the potential.

\begin{figure}
    \centering
    \includegraphics{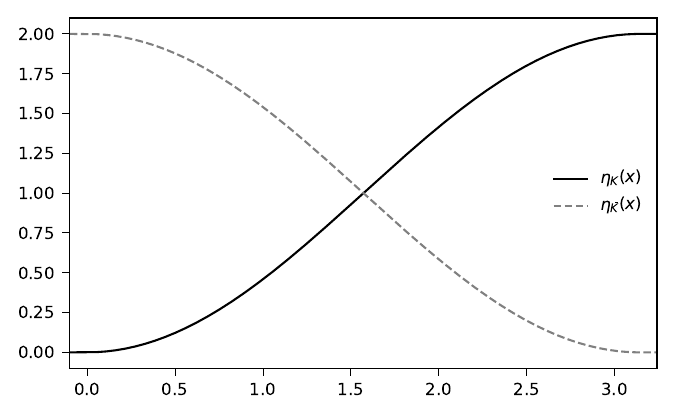}
    \caption{The kink $\eta_{(0,2)}(x)$ and the antikink $\eta_{(2,0)}(x)$ connecting the first two vacua $\eta=0$ and $\eta=2$.}
    \label{fig:kink}
\end{figure}

A basic kink connecting the first two vacua has the expression
\begin{equation*}
    \eta_K(x) =
    \begin{cases}
        0 & \text{ for } x < 0, \\
        1 - \cos x & \text{ for } 0 \le x \le \pi, \\
        2 & \text{ for } \pi < x.
    \end{cases}
\end{equation*}
An antikink connecting these vacua can be written simply as $\eta_{\bar{K}}=2-\eta_K(x)$.
We present the profile of these two solutions in figure~\ref{fig:kink}.
This pair of basic solutions can be used to construct kinks and antikinks connecting any pair of neighboring vacua by adding even integers to the expressions above in the form
\begin{align*}
    \eta_{(2k, 2k+2)}(x) &= \eta_{K}(x) + 2k, \\
    \eta_{(2k+2, 2k)}(x) &= \eta_{\bar{K}}(x) + 2k = -\eta_{K}(x) + 2(k+1)
\end{align*}
for $k \in \mathbb{Z}$.
A moving kink/antikink can be obtained through a Lorentz boost
\begin{equation*}
    \psi_{(2k, 2k\pm 2)}(t, x; v) = \eta_{(2k, 2k\pm 2)}(\gamma(x - v t))
\end{equation*}
where $\gamma = (1-v^2)^{-1/2}$.

The model defined by~\eqref{eq:action} has Hamiltonian density
\begin{equation*}
    \mathcal{H} = \frac{1}{2}(\partial_t\eta)^2 + \frac{1}{2}(\partial_x\eta)^2 + V(\eta)
\end{equation*}
which can be integrated in $x$ to give the energy.
In particular for kinks and antikinks at rest the energy is $\pi/2$, which we identify as the mass of the soliton.

\subsection{Collective coordinate approximation}

An important tool to describe the motion of solitons in field theory processes is the collective coordinate approximation.
For this we suppose the field time dependence can be described through a finite number of degrees of freedom $q(t) = (q^1(t), q^2(t), \ldots)$, such that $\eta(t, x) = \eta(x, q(t))$.
Starting from the original field theory action, the time evolution of the coordinates $q$ is determined by the Lagrangian
\begin{equation}
    L = \int dx \left[\frac{1}{2}(\partial_t\eta)^2-\frac{1}{2}(\partial_x\eta)^2-V(\eta)\right] = \frac{1}{2} g_{ij}(q) \, \dot{q}^i \dot{q}^j - U(q) \label{eq:cc-L}
\end{equation}
where we introduce a metric
\begin{equation*}
    g_{ij}(q) = \int dx \, \frac{\partial\eta}{\partial q^i} \frac{\partial\eta}{\partial q^j}
\end{equation*}
and a potential
\begin{equation*}
    U(q) = \int dx \left[ \frac{1}{2} (\partial_x \eta)^2 + V(\eta) \right]
\end{equation*}
in the space of the coordinates $q$.

The equations of motion for the Lagrangian in equation~\eqref{eq:cc-L} are
\begin{equation}
    g_{ij} \left(\ddot{q}^j + {\Gamma^j}_{mn} \dot{q}^m \dot{q}^n \right)+ \frac{\partial U}{\partial q^i} = 0
    \label{eq:eom-q}
\end{equation}
where
\begin{align*}
    {\Gamma^j}_{mn} &= \frac{1}{2} g^{jl} \left( \frac{\partial g_{nl}}{\partial q^m} + \frac{\partial g_{ml}}{\partial q^n} - \frac{\partial g_{mn}}{\partial q^l} \right) \\
    &= g^{jl} \int dx \, \frac{\partial\eta}{\partial q^l} \frac{\partial^2 \eta}{\partial q^m \,\partial q^n}
\end{align*}
are the Christoffel symbols and $g^{jl}$ is the inverse of the metric.
When the potential $U$ is a constant, equation~\eqref{eq:eom-q} is a geodesic equation.
Otherwise, the derivative of $U$ can be written as
\begin{equation*}
    \frac{\partial U}{\partial q^i} = \int dx \left[ \frac{\partial^2 \eta}{\partial q^i \, \partial x} \frac{\partial\eta}{\partial x} + V'(\eta) \frac{\partial \eta}{\partial q^i} \right].
\end{equation*}
In general, we can write the equations of motion as
\begin{equation}
    \ddot{q}^j \int dx \, \frac{\partial\eta}{\partial q^i} \frac{\partial\eta}{\partial q^j}
    + \dot{q}^j \dot{q}^k \int dx \, \frac{\partial\eta}{\partial q^i} \frac{\partial^2 \eta}{\partial q^j \,\partial q^k}
    + \int dx \left[ \frac{\partial^2 \eta}{\partial q^i \, \partial x} \frac{\partial\eta}{\partial x} + V'(\eta) \frac{\partial \eta}{\partial q^i} \right]
    = 0.
    \label{eq:cc-eq}
\end{equation}
Formally, the integrals above are to be taken over the entire $x$-axis, however, since we are dealing with compact solutions, the integration can be performed only on the support of $\eta$.
The more explicit formula of equation~\eqref{eq:cc-eq} is very useful for numerically solving for the time evolution of $q$ when the expression for the metric and the potential are too complicated.

Let us apply this formalism first to describe a single kink.
For this, we write the field as
\begin{equation*}
    \eta = \eta_{(2k, 2k \pm 2)}\left(x - a + \frac{\pi}{2}\right)
\end{equation*}
where $a$ represents the position of the kink center of momentum.
The metric only has one component
\begin{equation*}
    g(a) \equiv g_{aa}(a)
    = \int_{a-\pi/2}^{a+\pi/2} dx \, \left( \frac{\partial\eta}{\partial a} \right)^2
    = \frac{\pi}{2}
\end{equation*}
and the potential is a constant
\begin{equation*}
    U(a) = \frac{\pi}{2}.
\end{equation*}
Therefore, the kink dynamics is approximately that of a free particle of mass $\pi/2$.
Since this description did not take into account the Lorentz contraction of the kink support, we say that it is a non-relativistic description.

If we include a new coordinate $b$ to model the Lorentz contraction, such that the field is written as
\begin{equation*}
    \eta = \eta_{(2k, 2k \pm 2)}\left(b(x-a) + \frac{\pi}{2}\right)
\end{equation*}
we will have the diagonal metric with components
\begin{equation*}
    g_{aa}(a, b) = \frac{\pi  b}{2}, \quad
    g_{bb}(a, b) = \frac{\pi  \left(\pi ^2-6\right)}{24 b^3}, \quad
    g_{ab}(a, b) = 0
\end{equation*}
and the potential
\begin{equation*}
    U(a, b) = \frac{\pi  \left(1+b^2\right)}{4 b}.
\end{equation*}
The equations of motion for $a$ and $b$ are
\begin{gather*}
    \frac{d}{dt}\left( \frac{1}{2} \pi  \dot{a} b \right)  =0 \\
    \frac{d}{dt}\left[ \frac{\pi  \left(\pi ^2-6\right) \dot{b}}{24 b^3} \right] -\frac{1}{4} \pi  \dot{a}^2+\frac{\pi  \left(\pi ^2-6\right) \dot{b}^2}{16 b^4}-\frac{\pi  \left(b^2+1\right)}{4 b^2}+\frac{\pi }{2}  =0
\end{gather*}
which unsurprisingly have stationary solutions $\dot{a} = v$ and $b = (1-v^2)^{-1/2}$ for constant $v$.
This solution correspond to the Lorentz boosted expression of an exact kink solution.

Other non-stationary solutions can be used to describe dynamical process where the kink is perturbed from its exact form.
However, the scale factor is often insufficient to accurately describe the dynamics of perturbed kinks.
We must also include a perturbation to the shape of the kink profile.
Let us consider a basic kink with its center aligned with the origin, i.e.\ $a = 0$ and $b = 1$.
If add a perturbation $\chi$ such that $\eta(t, x) = \eta_K(x + \pi/2) + \chi(t, x)$, we obtain from the field equation for $0 < \eta < 2$ that
\begin{equation*}
    \partial_t^2 \eta_K - \partial_x^2 \eta_K + \sgn(\eta_K + \chi) - \eta_K + \partial_t^2 \chi - \partial_x^2 \chi - \chi = 0.
\end{equation*}
For $\chi$ small enough that $\sgn(\eta_K + \chi) = \sgn\eta_K$, we have
\begin{equation*}
    \partial_t^2 \chi - \partial_x^2 \chi - \chi = 0
\end{equation*}
which is a Klein-Gordon equation for $m^2 = - 1$.
This allows small perturbations of very general form.
Note that we do not have an obvious shape mode like the one in the $\phi^4$ model.

To allow the use of the collective coordinates approximation, we shall consider perturbations that respect the parity of the kink and that are non-trivial only over the kink-bulk.
Perturbations of this kind have Fourier expansion of the form
\begin{equation*}
    \chi(t, x) =
    \begin{cases}
        \sum_{n=1}^{\infty} c_n(t) \sin\left(2 n x \right) & \text{ if } -\frac{\pi}{2} < x < \frac{\pi}{2}, \\
        0 & \text{ otherwise}.
    \end{cases}
\end{equation*}
If impose that $\partial_x \eta$ must be continuous at $x = \pm \pi / 2$, we obtain the constraint
\begin{equation*}
    \sum_{n=1}^\infty (-1)^n n c_n(t) = 0. \label{eq:constraint}
\end{equation*}
We will restrict ourselves to use only the first two terms of the Fourier series.
This means that we have only one Fourier coefficient $c \equiv c_1 = 2 c_2$ that will be used as a collective coordinate.

This means that the internal mode of the kink is given by the function
\begin{equation}
    \chi(x) =
    \begin{cases}
        \sin(2x) + \frac{1}{2} \sin(4x) & \text{ if } -\frac{\pi}{2} < x < \frac{\pi}{2}, \\
        0 & \text{ otherwise}.
    \end{cases} \label{eq:internal-mode}
\end{equation}
Therefore, a more general expression to model the dynamics of a single kink is
\begin{equation*}
    \eta = \eta_K\left(b(x-a) + \frac{\pi}{2} \right) + c \, \chi\big(b(x-a) \big)
\end{equation*}
where we included the amplitude $c$ of the internal mode as an additional coordinate.
The procedure we use to obtain this internal mode is further justified by the phenomenology of our model.
It has been shown that the inclusion of the internal mode~\eqref{eq:internal-mode} significantly improves the description of deformed kinks~\cite{Hahne:2022wyl}.
As we shall see, this internal mode will also prove to be important in the description of the kink-antikink scattering.

\section{Kink-antikink scattering}
\label{sec:kink-antikink-scattering}

The compactness of the kinks allow us to construct exact solutions of the field equations, containing multiple solitons.
Consider a spatially symmetric configuration consisting of a kink-antikink pair, moving towards one another.
The kink connects the vacua $\eta = 0$ and $\eta = 2$, while the antikink connects $\eta = 2$ to $\eta = 0$, so that the whole configuration is topologically equivalent to the vacuum $\eta = 0$.
The kink comes from the left with velocity $v$ and the antikink comes from the right, with velocity $-v$.
Before they collide, the field is exactly described by a simple sum of the kink and antikink solutions.
Defining the time that the collision starts as $t = 0$, the initial condition for the time evolution of the scattering process is given by
\begin{align*}
    \eta(0, x) &= \psi_{(0, 2)}(0, x + \pi/\gamma; v) + \psi_{(2, 0)}(t, x; -v) - 2 \\
    \partial_t \eta(0, x) &= \partial_t\psi_{(0, 2)}(0, x + \pi/\gamma; v) + \partial_t\psi_{(2, 0)}(t, x; -v).
\end{align*}
At the moment of collision, the kink and antikink supports are, respectively, the intervals $-\pi/\gamma < x < 0$ and $0 < x < \pi / \gamma$.

\subsection{Numerical results}

\begin{figure}
    \includegraphics{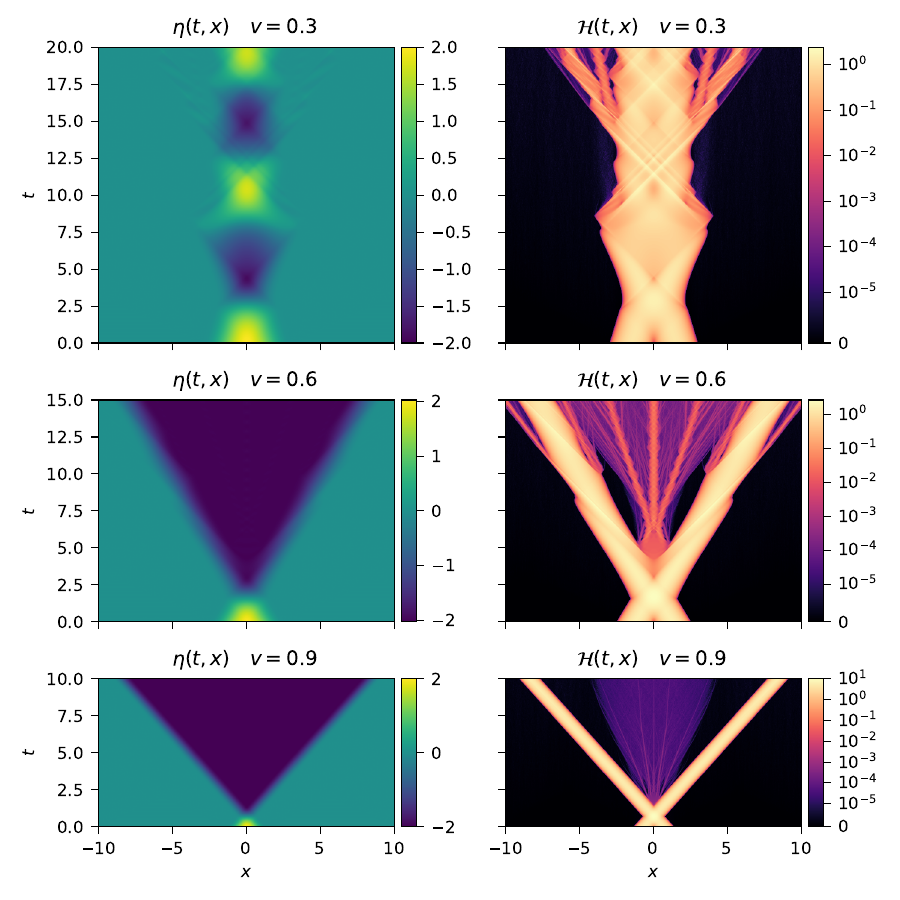}
    \caption{Field $\eta(t, x)$ and Hamiltonian density $\mathcal{H}(t, x)$ for kink-antikink scattering with velocity $v$.}
    \label{fig:kink_antikink/examples}
\end{figure}

We numerically evolved the field in time from the initial conditions above.
The time evolution of the field equation was done with the fourth-order Runge--Kutta method using Julia~\cite{Bezanson:2017} and the library DifferentialEquations.jl~\cite{Rackauckas:2017}.
Different time steps $\Delta t$ ranging from $10^{-4}$ to $10^{-3}$ where used and yielded consistent results.
The spatial derivatives were discretized by finite differences up to second-order.
The spatial grid was taken with separation $\Delta x = 10 \Delta t$.
Some example plots of the field and the Hamiltonian density are presented in figure~\ref{fig:kink_antikink/examples}.

For small scattering velocities bellow some critical velocity $v_c$, the kink and antikink collision leads to the creation of field configuration that oscillates between the vacua $\eta = \pm 2$.
For larger velocities, the solitons are able to pass through one another, however, the vacuum in the region between them changes from $\eta=2$ to $\eta=-2$.
Since the model is not integrable, there is emission of radiation, which can be seen in the Hamiltonian density plots.
This radiation is more pronounced for smaller scattering velocities, and it decreases as the velocity increases.

\begin{figure}
    \includegraphics{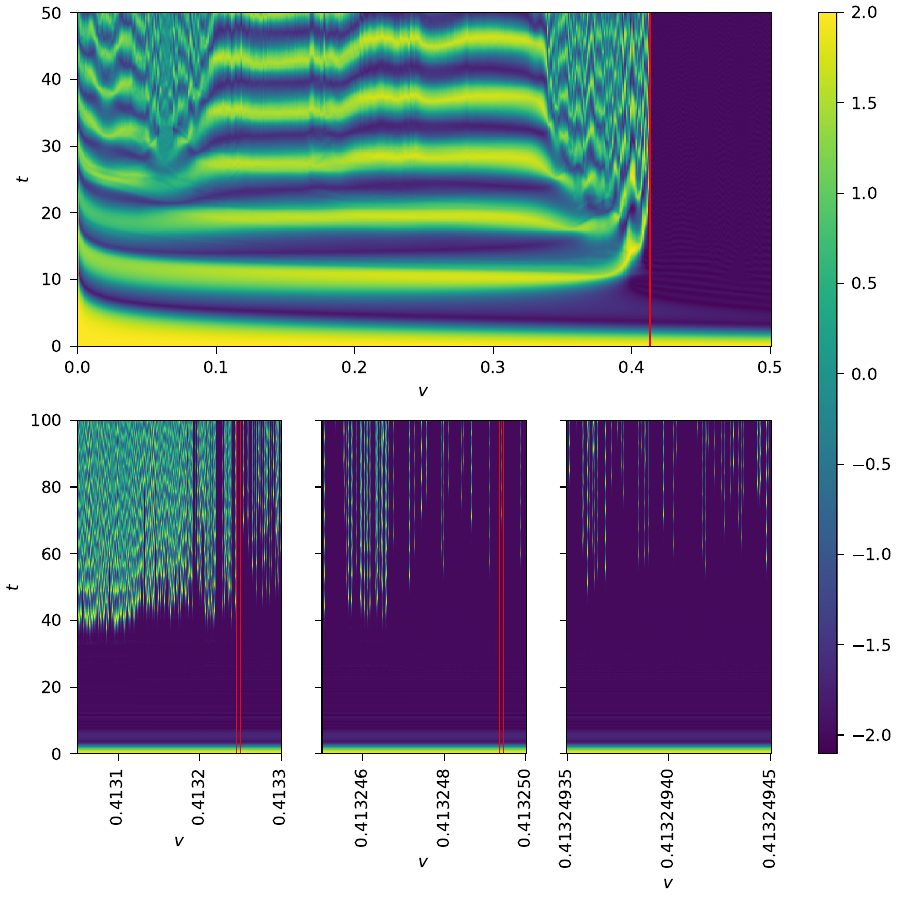}
    \caption{Field value $\eta(t, 0)$ as a function of time $t$ and velocity $v$ for a kink-antikink scattering. The red lines correspond to the range of $v$ of the following panel.}
    \label{fig:kink-antikink/middle}
\end{figure}

We can identify if the kink and the antikink pass through each other or not by looking at the field value at $x=0$ as a function of time.
If the field $\eta(t, 0)$ oscillates, the scattering resulted in the annihilation of the solitons.
If $\eta(t, 0)$ reaches a vacuum different from zero and remains there, a pair kink-antikink emerged from the collision.
We present the field values $\eta(t, 0)$ at the middle point $x=0$ as a function of time $t$ and scattering velocity $v$ in figure~\ref{fig:kink-antikink/middle}.

There is a critical velocity $v_c \approx 0.413$ such that, in general, if $v \lesssim v_c$ the solitons annihilate, and if $v \gtrsim v_c$ a pair kink-antikink emerges.
One particularity of our results is that there is no window of velocities for which the pair oscillates a few times and then separates.
However, this does not mean that the scattering behavior is monotonic in $v$.
There are two windows, $ 0.05 \lesssim v \lesssim 0.1 $ and $0.34 \lesssim v \lesssim v_c$, where the field behavior depends on $v$ rather chaotically.

Also, the transition between annihilation and escape of the pair happens in a fractal manner.
This can be seen in the lower panels of figure~\ref{fig:kink-antikink/middle}, where we zoom into smaller windows of $v$ close to the critical velocity.
We checked changes of velocity as small $10^{-8}$ and observed alternating cases of annihilation and escape for velocities close to $v_c$.
This fractal behavior was also reported in different models with compact solutions~\cite{Bazeia:2019tgt}.

\begin{figure}
    \includegraphics{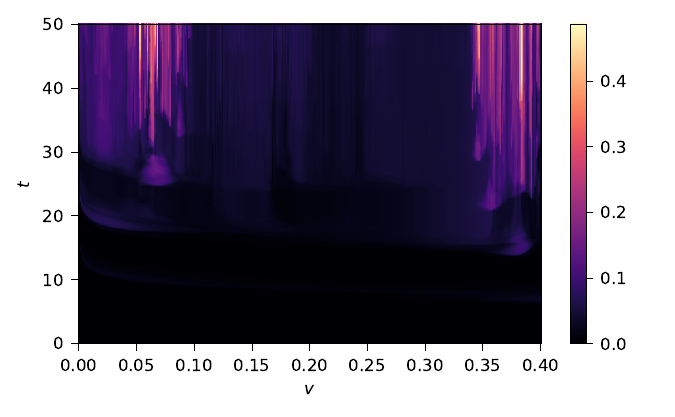}
    \caption{Fraction of total energy carried away by radiation during a kink-antikink scattering for scattering velocities $v<v_c$, i.e.~when no kink or antikink emerge from the collision.}
    \label{fig:kink_antikink/radiation_energy}
\end{figure}

The more chaotic behavior is strongly linked to the emission of radiation.
We estimate the radiation carried away by radiation for $v<v_c$ as the energy outside the region $ -\pi < x < \pi$.
This radiation energy can be written as the integral
\begin{equation*}
    E_\text{rad}
    = \int_{-\infty}^{-\pi} dx \, \mathcal{H} + \int_{\pi}^{\infty} dx \, \mathcal{H}= 2 \int_{\pi}^{\infty} dx \, \mathcal{H}
\end{equation*}
that can be calculated from the numerical data.
In figure~\ref{fig:kink_antikink/radiation_energy} we present the ratio $E_\text{rad} / (\gamma \pi)$, where $\gamma \pi$ is the total energy of the scattering process.
The regions with the most emission of radiation match the regions of chaotic behavior in figure~\ref{fig:kink-antikink/middle}.

It has been previously reported that the radiation in models with non-analytical potentials is dominated by compact oscillons~\cite{Klimas:2018woi}.
This oscillons can be directly emitted from a perturbed kink~\cite{Hahne:2022wyl}.
The direct emission of compact oscillons occurs in the form of jets leaving the kink in the Hamiltonian density plots of figure~\ref{fig:kink_antikink/examples}.

\begin{figure}
    \includegraphics{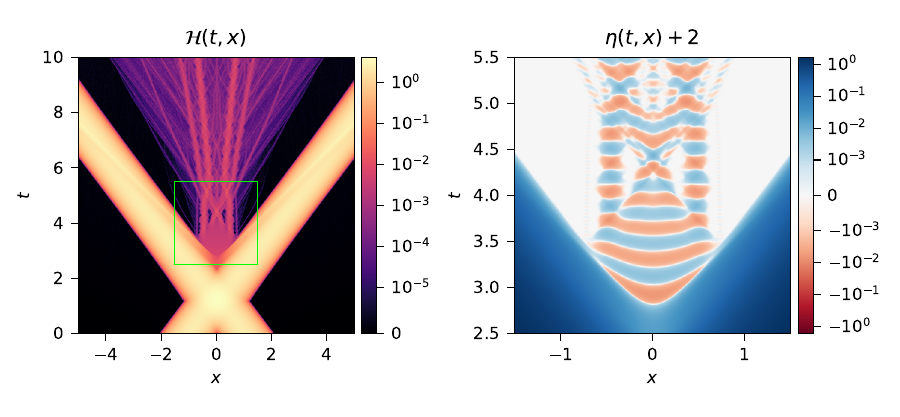}
    \caption{Kink-antikink scattering for $v=0.75$. Left: Hamiltonian density. Right: zoom into the difference between field value and vacuum $-2$, i.e.~$\eta(t,x) + 2$, for the region inside the green rectangle on the right panel.}
    \label{fig:kink_antikink/zoom}
\end{figure}

Other possibility occurs when the kink-antikink pair escapes.
In this case, in the region between the two topological defects, emerges a diamond shaped configuration in the spacetime diagram, see figure~\ref{fig:kink_antikink/zoom}.
For this configuration, the field reaches vacuum values at hyperbolas in spacetime.
This behavior is characteristic of compact shockwaves of the signum-Gordon model.
Compact shockwaves are exact solutions of the signum-Gordon model~\cite{Arodz:2005bc}, which correspond to the limit of small amplitudes around the vacua of our model.
These solutions can be obtained from initial data in the shape of a Dirac delta function.
As a consequence of sharp initial conditions, the wavefront in each border of the support is discontinuous, making the whole shockwave have infinite energy.
This makes the shockwaves borders correspond to an energy reservoir, allowing the shockwave support to grow with the speed of light.
It has been shown that when the initial data of the shockwave is replaced by a smooth delta-like function, the shockwave has finite energy and eventually decays into a cascade of oscillons when the wavefront energy reservoir is depleted~\cite{Hahne:2019odw}.
This decay makes the shockwave present itself with a diamond shape in spacetime, like we see in our simulations.
The same kind of shockwave appears in the scattering of compact oscillons in the signum-Gordon model~\cite{Hahne:2019ela}.

Note that the scattering process does not lead to the creation of new kink-antikink pairs.
The pair creation could happen for cases with energy $E = \gamma \pi \geq 2\pi$, i.e.\ $v \geq \sqrt{3} / 2 \approx 0.866$.
However, this does not happen in our simulations.
The energy of the collision does not transfer from the solitons translational mode to the field amplitude in enough quantities to make the field reach different vacua in the region between the original pair.

\subsection{Collective coordinate approximation}

\begin{figure*}
    \includegraphics{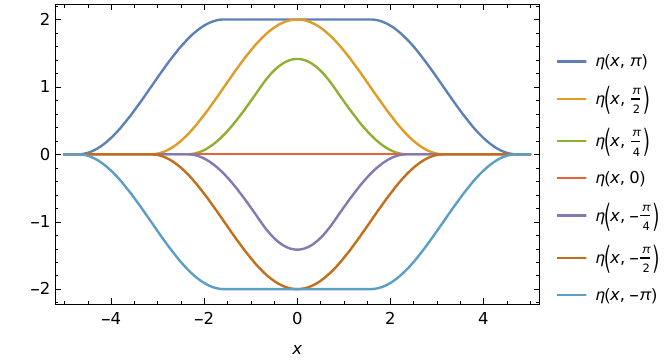}
    \caption{Field profile for a superposition of kink and antikink.}
    \label{fig:kink_antikink/non-rel-eta}
\end{figure*}

In this section, we will approach the problem of a kink-antikink scattering using the collective coordinate approximation.
We begin by assuming that during the whole process the field can be written as a symmetric superposition of a kink and an antikink, in the form
\begin{equation}
    \eta(x, a) = \eta_K\left(x + a + \frac{\pi}{2}\right) - \eta_K\left(x - a + \frac{\pi}{2}\right) \label{eq:kak-nonrel-eta}
\end{equation}
where $a$ is the center of momentum position of the antikink.
We do not need to introduce a different coordinate for the kink position because of the symmetry of the problem.
The profile of the field can be seen in figure~\ref{fig:kink_antikink/non-rel-eta}.

\begin{figure}
    \includegraphics{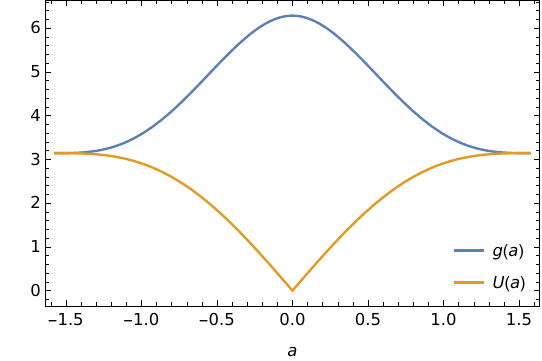}
    \caption{Metric and potential for a collective coordinate model of a kink-antikink collision with only a positional mode.}
    \label{fig:kink_antikink/non-rel-metric-and-potential}
\end{figure}

Note that the expression for the field is an odd function in $a$.
Since the metric and potential are independent of the sign of the field, we know that they will be a function of $\abs{a}$ only.
Since the field expressions are ultimately given by piecewise formulas, we must take care of each of the following cases.
\begin{itemize}
    \item If $\abs{a} \geq \pi/2$, the kink and antikink supports do not overlap, and in particular for $\dot{a}=0$, the configuration corresponds to an exact solution of two static solitons.
    \item If $\abs{a} < \pi / 2$, there is a region $\abs{a}-\pi/2 < x < -\abs{a}+\pi/2$ where both solitons overlap.
    Since the solitons have compact support, this is the only situation when the solitons interact.
    This is fundamentally different from models like the $\phi^4$ model, where there is interaction even at larges distances between the centers of momentum.
\end{itemize}
Doing the necessary integrations, taking into account the cases above, we obtain the following expressions for the metric and the potential:
\begin{align*}
    g(a) &=
    \begin{cases}
        \pi & \text{ for } \abs{a} \geq \frac{\pi}{2}, \\
        \pi+\sin (2 \abs{a})+(\pi - 2 \abs{a}) \cos (2 a) & \text{ for } \abs{a} < \frac{\pi}{2},
    \end{cases}\\
    U(a) &=
    \begin{cases}
        \pi & \text{ for } \abs{a} \geq \frac{\pi}{2}, \\
        2 \abs{a}+\sin (2 \abs{a}) & \text{ for } \abs{a} < \frac{\pi}{2}.
    \end{cases}
\end{align*}
The shape of these functions in the non-trivial interval $\abs{a} < \pi/2$ is shown in figure~\ref{fig:kink_antikink/non-rel-metric-and-potential}.

The time evolution of $a$ can be determined either by an Euler-Lagrange equation or from the energy conservation condition
\begin{equation*}
    \frac{1}{2} g(a) \dot{a}^2 + U(a) = E.
\end{equation*}
From the latter, we can obtain $a(t)$ implicitly from the integral
\begin{equation*}
    \int_{a(0)}^{a(t)} da \sqrt{\frac{g(a)}{2(E - U(a))}} = \pm t
\end{equation*}
with the sign to be determined from the sign of $\dot{a}(0)$.
For the scattering process we have
\begin{equation*}
    E = \frac{\pi}{2} v^2 + \pi.
\end{equation*}
Note that $E > U(a)$ for every $a$, so the trajectory of $a$ does not have any point of return during the collision.
This means, that the collective coordinate approximation predicts that the kink and antikink will always pass through each other.
This is in contradiction with the simulation results.

We can interpret this contradiction as an indication that some mode of the field is activated during the collision that robs the position degree of freedom of some of its energy.
One could imagine that the Derrick mode related to the Lorentz contraction could be responsible for this.
We could write an expression for the field of the form
\begin{equation*}
    \eta(x, a, b) =  \eta_K\left(b(x + a) + \frac{\pi}{2}\right) - \eta_K\left(b(x - a) + \frac{\pi}{2}\right)
\end{equation*}
where the scale factor $b$ takes care of the Lorentz contraction.
However, note that
\begin{equation*}
    \frac{\partial\eta(x, a, b)}{\partial b} = (x + a) \, \eta_K'\left(b(x + a) + \frac{\pi}{2}\right) - (x - a) \, \eta_K'\left(b(x - a) + \frac{\pi}{2}\right)
\end{equation*}
and in particular for $a = 0$, the two terms cancel out and $\partial_b \eta = 0$.
Therefore, an entire row and column of the metric matrix is null, and so is the determinant of the metric.
At the same time, the potential has the form
\begin{equation*}
    U(a, b) =
    \begin{cases}
        \frac{\pi  \left(b^2+1\right)}{2 b} & \text{ for } \abs{a} \geq \frac{\pi}{2b}, \\
        -\frac{\left(b^2-3\right) \sin (2 b | a| )+\left(b^2-1\right) (\pi -2 b | a| ) \cos (2 b | a| )-4 b | a| -\pi  b^2+\pi }{2 b} & \text{ for } \abs{a} < \frac{\pi}{2b}
    \end{cases}
\end{equation*}
which has a minimum at $a=0$.
Therefore, nothing forbids the evolution of the coordinates of reaching the point $a=0$.
However, at this point the metric is degenerate and can not be inverted to solve the equations of motion.
A similar problem appears in different models, such as the $\phi^4$ model, where it can be solved by adding the relativistic contribution perturbatively~\cite{Adam:2021gat}.
However, this approach can not be applied to the scattering of compactons because we can not truncate the Taylor series of $\eta$ around $b=1$ without losing the continuity of $\partial_x \eta$ at the kink support boundaries.

However, there are other possible factors that can explain the existence of points of return during the collision.
The emission of radiation, for example, spends part of the collision energy.
Another possibility is the activation of some internal mode of the solitons, like the one presented in equation~\eqref{eq:internal-mode}.
Therefore, a tentative expression for the field is the following:
\begin{equation*}
    \eta(x, a, c) = \eta_K\left(x + a + \frac{\pi}{2}\right) - \eta_K\left(x - a + \frac{\pi}{2}\right) + c \big[\chi(x + a) - \chi(x - a) \big].
\end{equation*}
However, once again we have a problem.
Since the internal modes cancel out when the solitons superpose we have $\partial_c \eta = 0$ for $a=0$.
A similar problem happens in the $\phi^4$ model when the shape-mode is included in the description of kink-antikink scattering, and it is referred as a null-vector problem~\cite{Caputo:1991cv,Takyi:2016tnc}.
In the $\phi^4$ case, this problem has been solved by a change of variables $c \to c / f(a)$, where $f(a)$ is a linear function in $a$ for small $a$~\cite{Manton:2020onl,Manton:2021ipk}.
Since, for small values the $a$, the superposition of internal modes has the expansion
\begin{align*}
    \chi (a+x)-\chi (x-a) = 4 ac \, (\cos (2 x)+\cos (4 x))+\mathcal{O}\left(a^3\right)
\end{align*}
we can adopt this same technique here, and write the field as
\begin{equation*}
    \eta(x, a, c) = \eta_K\left(x + a + \frac{\pi}{2}\right) - \eta_K\left(x - a + \frac{\pi}{2}\right) + \frac{c}{f(a)} \big[\chi(x + a) - \chi(x - a) \big].
\end{equation*}
The precise formula for $f(a)$ is not important, as long it has expansion $f(a) \sim a$ for small $a$.
In particular, we chose
\begin{equation*}
    f(a) =
    \begin{cases}
        \sin(a) & \text{ if } -\frac{\pi}{2} < a < \frac{\pi}{2}, \\
        \sgn(a) &\text{ otherwise},
    \end{cases}
\end{equation*}
so that the null-vector problem is properly taken care of when the solitons superpose, but the interpretation of $c$ as a mode amplitude remains when the solitons are not interacting.

\begin{figure}
    \includegraphics{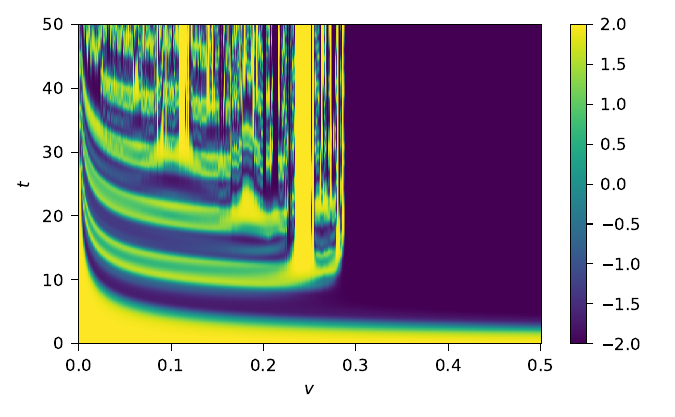}
    \caption{Collective coordinates prediction for field value $\eta(x=0, a(t), c(t))$ as a function of time $t$ and velocity $v$ for a kink-antikink scattering. For this result, we included the position and internal mode into the description.}
    \label{fig:kink-antikink/middle_point_cc}
\end{figure}

The metric and the potential in this case have very long expressions, which we will not write here.
In particular, the formula of the potential will change whether there are regions where the internal mode changes the signal of $\eta + 2k$, for $k \in \mathbb{Z}$.
Taking care of all this cases analytically can become prohibitively difficult.
However, the time evolution of $a$ and $c$ can be obtained numerically in a straightforward manner by calculating the integrals and solving the equation~\eqref{eq:cc-eq} with a symplectic method.
We summarized the results in a plot of the field in the middle point $x = 0$ as a function of time and velocity, similar to what was done for the simulations.
This plot is presented in figure~\ref{fig:kink-antikink/middle_point_cc}.

The behavior of $\eta$ at $x=0$ indicates that this collective coordinate model with $a$ and $c$ captures the main features of the scattering process.
For small enough velocities, the pair forms an oscillating configuration, being able to escape only for larger velocities.
However, the critical velocity is about $0.3$, which is smaller than value from the full field theory simulations.
This can be attributed, at least partially, to the emission of radiation.
When the configuration irradiates, it loses energy and needs a larger initial energy to be able to escape the bound state.
It is also possible that there are further internal modes that are not accounted.
For example, due to the difficulties mentioned above, we can not take the Lorentz contraction into account, even tough the pair annihilation is clearly a relativistic process.

Another visible difference is the existence of windows of velocity where the kink and antikink are able to escape to the same side from where they came.
This corresponds to the cases where the field at $x=0$ returns and remains at $\eta=2$.
This is not something that happens in the full field dynamics.
Once again, possible explanations for the difference radiation and unaccounted modes.

\section{Kink-kink scattering}
\label{sec:kink-kink-scattering}

We will now consider the scattering process of two compact kinks.
In particular, let us consider a kink connecting the vacua $\eta=-2$ and $\eta=0$ moving from the right with velocity $v$, while a kink connecting the vacua $\eta=0$ and $\eta=2$ moves from the left with velocity $-v$.
The initial data is of the form
\begin{align*}
    \eta(0, x) &= \psi_{(-2,0)}(0, x + \pi/\gamma; v) + \psi_{(0,2)}(0, x; -v) \\
    \partial_t \eta(0, x) &= \partial_t\psi_{(-2,0)}(0, x + \pi/\gamma; v) + \partial_t\psi_{(0,2)}(0, x; -v).
\end{align*}
At the initial instant, the kink $\psi_{(-2,0)}$ is located at $x \in [-\pi/\gamma, 0]$ and its moving to the right, while the kink $\psi_{(0,2)}$ is located at $x \in [0, \pi/\gamma]$ and its moving to the left.
Each kink has energy $\gamma \pi/2$, such that the total energy of this configuration is $\gamma \pi$.

While the form of the initial condition is somewhat similar to that of the kink-antikink scattering, this is a very different process because the kink-kink pair has a non-zero topological charge.
Because topological charges are a conserved quantity under continuous mappings, including time evolution, the pair does not annihilate.
Furthermore, the initial configuration is an odd function of $x$.
Since parity is a symmetry of the equations of motion, the field at the middle point between the two kinks must remain static, i.e.~$\eta(t, 0) = 0$.
This severely constraints any form of radiation to be formed during the scattering.
Therefore, we expect the collision to only cause an exchange of momentum with very small amounts of radiation, possibly none.

\begin{figure}
    \includegraphics{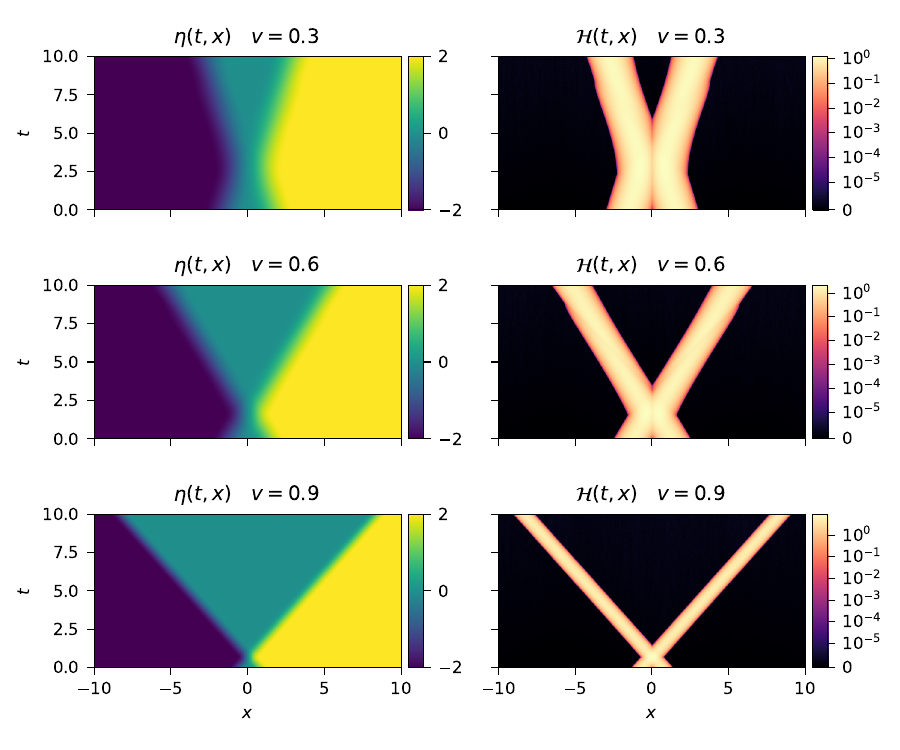}
    \caption{Field $\eta(t, x)$ and Hamiltonian density $\mathcal{H}(t, x)$ for kink-kink scattering with velocity $v$.}
    \label{fig:kink-kink/examples}
\end{figure}

We can confirm our expectations by simulating numerically the time evolution of the field.
We present some examples of the simulation results in figure~\ref{fig:kink-kink/examples} as color maps in a spacetime diagram.
The kinks just bounce against each other, with no emission of radiation.
Note that the Hamiltonian density plot is logarithmic, so even small amounts of radiation would be visible.

\begin{figure*}
    \includegraphics{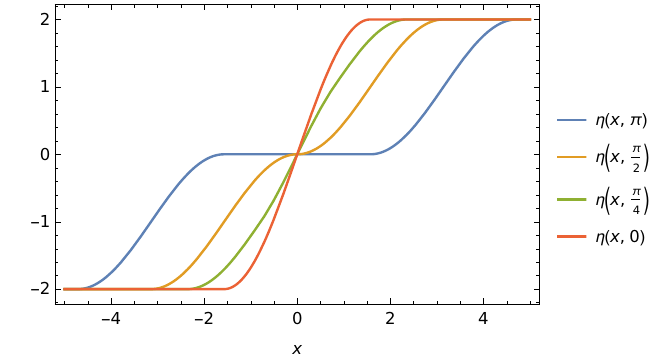}
    \caption{Field profile for a superposition of two kinks.}
    \label{fig:kink_kink/non-rel-eta}
\end{figure*}

The rather simple behavior of the scattering process makes us very tempted to try to describe it using collective coordinates.
We will consider the following four possible sets of coordinates.
\begin{itemize}
    \item {Non-relativistic.}
    Case with only the positional mode.
    In this case the field is given by the expression simple superposition of two kinks.
    The only coordinate is the center of momentum position of the kink on the right.
    Therefore, the field has the form
    \begin{equation*}
        \eta(x, a) = \eta_{K}\left(x + a + \frac{\pi}{2}\right) + \eta_{K}\left(x - a + \frac{\pi}{2}\right) - 2
    \end{equation*}
    where the constant $-2$ is included to make the kinks connect the vacua $-2$, $0$, and $2$.
    Some examples of field profiles are shown in figure~\ref{fig:kink_kink/non-rel-eta}.
    \item {Non-relativistic with internal mode.}
    Case when we include the internal mode amplitude $c$ as a coordinate.
    The field is then given by
    \begin{equation*}
        \eta(x, a, c) = \eta_{K}\left(x + a + \frac{\pi}{2}\right) + \eta_{K}\left(x - a + \frac{\pi}{2}\right) - 2 + c \left[ \chi(x+a) + \chi(x-a) \right].
    \end{equation*}
    Note that, since the internal modes do not cancel for $a = 0$ in the kink-kink case, there is no need to regularize the amplitude $c$ like before.
    \item {Relativistic.}
    Case when we include a scale factor $b$ to model the Lorentz contraction.
    The field is written as
    \begin{equation*}
        \eta(x, a, b) = \eta_{K}\left(b(x + a) + \frac{\pi}{2}\right) + \eta_{K}\left(b(x - a) + \frac{\pi}{2}\right) - 2.
    \end{equation*}
    \item {Relativistic with internal mode.}
    Case when we include both the scale factor $b$ and the internal mode amplitude $c$, so that the field is
    \begin{equation*}
        \eta(x, a, b, c) = \eta_{K}\left(b(x + a) + \frac{\pi}{2}\right) + \eta_{K}\left(b(x - a) + \frac{\pi}{2}\right) - 2 + c \left[ \chi(b(x+a)) + \chi(b(x-a)) \right].
    \end{equation*}
\end{itemize}

Unfortunately, the kink-kink description has a null-vector problem with the positional mode $a$.
For all the four cases above, the derivative $\partial_a \eta$ is null for $a=0$.
For example, for the non-relativistic case the derivative is
\begin{equation*}
    \frac{\partial\eta(x, a)}{\partial a} = \eta_{K}'\left(x + a + \frac{\pi}{2}\right) - \eta_{K}'\left(x - a + \frac{\pi}{2}\right)
\end{equation*}
and the two terms cancel when the kinks overlap completely.
The same happens for all the four sets of coordinates, because they all include the positional mode similarly.

This null-vector problem also appears at the sine-Gordon model, where it was solved by allowing the $a$ coordinate to have imaginary values~\cite{Adam:2021gat}.
This was possible there because the specific formula for the field as a function of $a$ remains real even if $a$ is imaginary.
However, in our model $a$ also appears in the definition of the kink support, so we can not allow it to have imaginary values.
Similar to the case of relativistic kink-antikink scattering, the compactness of the kink support restricts our ability to use known methods of other models in a straightforward manner.
This highlights how models with non-analytical potentials are rather unique.
Even though they share several characteristics with more commonly studied models, they also come with several subtleties.

\begin{figure}
    \includegraphics{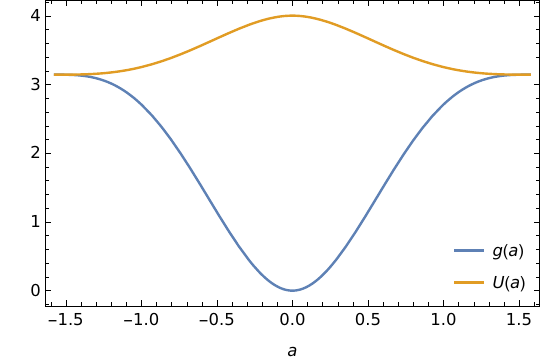}
    \caption{Metric and potential for a collective coordinate model of a kink-kink collision with only a positional mode.}
    \label{fig:kink_kink/non-rel-metric-and-potential}
\end{figure}

However, this does not mean that the collective coordinate approximation is useless in this case.
To illustrate this, let us calculate the expressions for the metric and the potential in the non-relativistic case.
By taking care of the different cases $\abs{a} > \pi / 2$ and $\abs{a} < \pi/2$, in the same fashion as for the kink-antikink scattering, we obtain
\begin{align*}
    U(a) &=
    \begin{cases}
        \pi & \text{ for } \abs{a} \geq \frac{\pi}{2}, \\
        2\abs{a}-\sin (2 \abs{a})+4 \cos (a) & \text{ for } \abs{a} < \frac{\pi}{2},
    \end{cases}\\
    g(a) &=
    \begin{cases}
        \pi & \text{ for } \abs{a} \geq \frac{\pi}{2}, \\
        \pi-\sin (2 \abs{a})-(\pi -2 \abs{a})\cos(2a) & \text{ for } \abs{a} < \frac{\pi}{2}.
    \end{cases}
\end{align*}
The shape of these functions is presented in figure~\ref{fig:kink_kink/non-rel-metric-and-potential}.

Note that the potential has the form a barrier for $\abs{a} < \pi/2$, differently from the kink-antikink case.
If the potential barrier at $a=0$ was infinite, the null-vector problem would be completely avoided, because the coordinate $a$ would never reach this point and the metric would always remain non-degenerate.
But, since $U(0) = 4$ the maximum of $U(a)$ is accessible if the total energy $E$ is high enough.
In the non-relativistic description the energy of the scattering is $E = \pi + \pi v^2 / 2$, therefore the point $a=0$ is accessible when $E=4$, i.e.\ when $v = \sqrt{8/\pi-2} \approx 0.739$.
Beyond this velocity, the time evolution for $a$ can not be obtained from the equations of motion.
Similar considerations are valid for the relativistic case, and both the cases with the internal mode.

\begin{figure}
    \centering
    \includegraphics{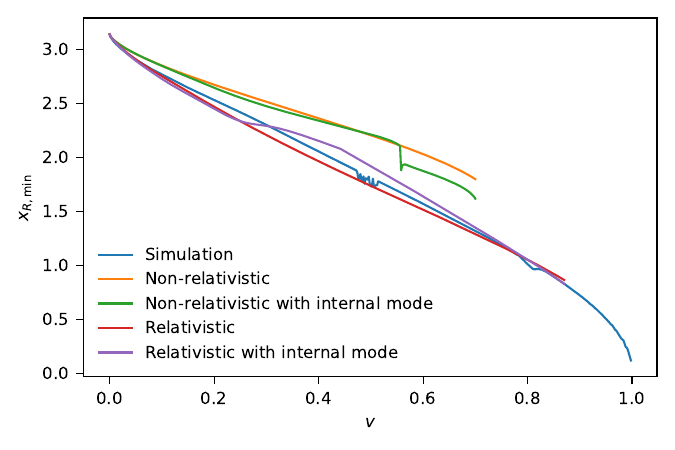}
    \caption{Values of $x_{R,\text{min}}$ calculated from the simulations and predicted by collective coordinates.}
    \label{fig:closest_approximation}
\end{figure}

These collective coordinate descriptions are not metrically complete.
However, for velocities small enough as to not reach to potential peak at $a=0$, we can still use them.
We would like to test the predictions from the collective coordinate approximation for such cases.
A way of doing this systematically it to calculate the closest approximation between the two kinks during the scattering and compare the results to the simulations.
Since the plots in figure~\ref{fig:kink-kink/examples} indicate that the kinks bounce against one another, the function $x_R(t)$, which represents the trajectory of the right border of the kink connecting $\eta=0$ to $\eta=2$, has a global minimum $x_{R,\text{min}}$.
The simulation values of $x_{R}(t)$ can be obtained by finding the smallest value of $x_R$ to satisfy $ |\eta(t, x_R) - 2| < 10^{-7}$.
This minimum can also be found for the collective coordinate approximation by minimizing the functions
\begin{equation*}
    x_{R}(t) = a(t) + \frac{\pi}{2}
\end{equation*}
for the non-relativistic cases, and
\begin{equation*}
    x_{R}(t) = a(t) + \frac{\pi}{2b(t)}
\end{equation*}
for the relativistic cases.
The values of $x_{R,\text{min}}$ for different scattering velocities $v$ is presented in figure~\ref{fig:closest_approximation}, where we show the results from the simulations and from solving the collective coordinate equations of motion~\eqref{eq:cc-eq}.

\begin{figure}
    \centering
    \includegraphics{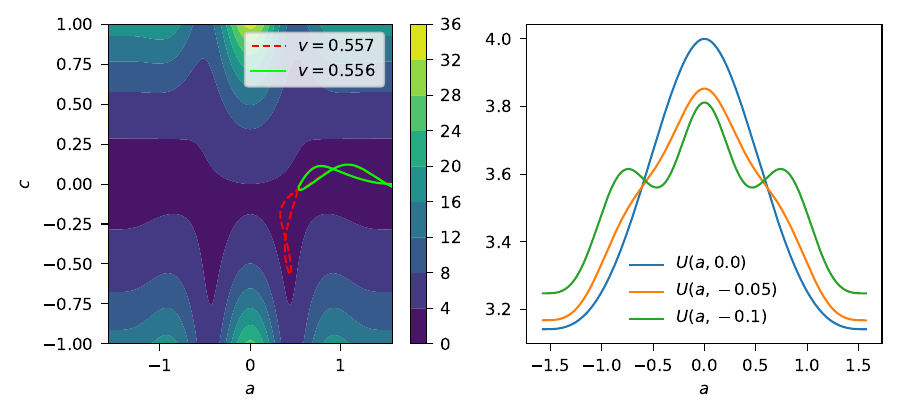}
    \caption{On the left, effective potential $U(a, c)$ for collective coordinate model of kink-kink scattering, including translational and internal modes, with lines for two solutions of the equations of motion. On the right, profile of the potential for fixed values of $c$.}
    \label{fig:kink_kink/non_rel_mode_potential}
\end{figure}

A curious feature of the predicted values of $x_{R,\text{min}}$ by the model non-relativistic with internal mode is the discontinuity at $v\approx 0.557$.
This can be explained by noticing that for higher values of $v$, the internal mode amplitude $c$ reaches values where the potential has new local minima.
Note in figure~\ref{fig:kink_kink/non_rel_mode_potential} that the potential $U(a, c)$ has something akin to a phase transition for $c\approx 0.05$.
This reduces the energy required for the kinks getting closer together, i.e.\ smaller values of $a$, explaining the jump in the curve for predicted $x_{R,\text{min}}$.
However, this feature is absent in the simulations, whose results are better described the relativistic model.
In the relativistic model, the inclusion of the internal mode does not significantly improve the predictions.

\section{Conclusions}
\label{sec:conclusions}

In this work we analyzed the interaction between compact topological defects.
First we looked at the kink-antikink scattering problem using numerical simulation.
The simulation results can be classified in two main classes.
In one class of scatterings, the kink-antikink pair forms an oscillating configuration between $\eta=-2$ and $\eta=2$ which emits excess energy in the form of radiation.
The second class correspond to the cases when the kink-antikink pair emerges from the collision.
In these cases, a shockwave is formed between the pair.
The shockwave eventually decays into a cascade of compact oscillons, similarly to what was reported in the collision of compact oscillons of the signum-Gordon model.
The higher scattering velocities display smaller shockwaves and radiation.

We would like to highlight that we found no evidence of cases where the field oscillates a few times, and then the pair escapes.
These ``multi-bounce'' cases are a key characteristic of kink-antikink scattering in the $\phi^4$ model, but are absent here.
Another interesting feature is the fractal behavior close to the critical velocity $v_c$.
Although for scattering velocities $v$ sufficiently larger than $v_c$, the kink-antikink pair always escapes, the escape and no-escape cases alternate in a fractal manner for $v$ close to $v_c$.
This seems to be a more general feature of compacton-anticompacton collisions.

We were able to use a collective coordinate model to describe some features of the kink-antikink collision, like the oscillating behavior for $v<v_c$.
The coordinates considered were a positional mode and the amplitude of an internal mode.
However, we were not able to precisely predict the value of $v_c$ using this model.
A more precise description may be hampered by the presence of high amounts of radiation.
The simulations show that cases with more radiation emission are the ones with more chaotic field behavior.

In the kink-kink case, we observed that the pair collides elastically with no emission of radiation.
This absence of radiation can be attributed to strong constraints placed by the parity of the initial data.
The collective coordinate model for kink-kink collisions was shown to exhibit a null-vector problem.
However, we were still able to obtain predictions for the position of the closest approximation of the pair in the range of velocities were the null-vector problem can be avoided.

Models with non-analytic potentials have several similarities to more usual field theory models, since they have standard kinetic terms, leading to hyperbolic differential equations for the fields.
However, the V-shape of the potential introduces sufficient difficulties to prevent us from adapting proven techniques from other models.
This has prevented us from solving the null-vector problems using the procedures from other models, specially the Klein-Gordon $\phi^4$ and the sine-Gordon models.
The difficulties arise from the fact that the field expressions are given by piecewise functions.
A collective coordinate description of any dynamical process that involves changes in the solitons size and or position inevitably leads to a dependency of the support on the collective coordinates itself.
This severely limits the ability of performing Taylor expansions or analytical continuations of the field on the collective coordinates.

The V-shaped potentials, unlike the ordinary potential, cannot be linearized around their minima.
This makes it difficult to formulate the eigenvalue problem for small kink perturbations.
Even though the existence of translational symmetry is obvious, the usual calculus of the translational zero mode is not straightforward.
Furthermore, the Taylor expansion of a kink function around its center is frequently problematic for compactons since it spoils the continuity of a compacton first derivative.
Unlike in popular models, such as the $\phi^4$ model, the first and higher Derrick modes cannot be obtained simply by series expansion.
It became clear that description of compact kinks dynamics require some further development of mathematical tools that may better grasp the key properties of such compactons.

\begin{acknowledgments}
We thank A.~Wereszczy\'nski for providing helpful comments.
F.~M.~Hahne is supported by CNPq--Brazil.
\end{acknowledgments}

\bibliography{bibliography}

\begin{thebibliography}{25}%
\makeatletter
\providecommand \@ifxundefined [1]{%
 \@ifx{#1\undefined}
}%
\providecommand \@ifnum [1]{%
 \ifnum #1\expandafter \@firstoftwo
 \else \expandafter \@secondoftwo
 \fi
}%
\providecommand \@ifx [1]{%
 \ifx #1\expandafter \@firstoftwo
 \else \expandafter \@secondoftwo
 \fi
}%
\providecommand \natexlab [1]{#1}%
\providecommand \enquote  [1]{``#1''}%
\providecommand \bibnamefont  [1]{#1}%
\providecommand \bibfnamefont [1]{#1}%
\providecommand \citenamefont [1]{#1}%
\providecommand \href@noop [0]{\@secondoftwo}%
\providecommand \href [0]{\begingroup \@sanitize@url \@href}%
\providecommand \@href[1]{\@@startlink{#1}\@@href}%
\providecommand \@@href[1]{\endgroup#1\@@endlink}%
\providecommand \@sanitize@url [0]{\catcode `\\12\catcode `\$12\catcode
  `\&12\catcode `\#12\catcode `\^12\catcode `\_12\catcode `\%12\relax}%
\providecommand \@@startlink[1]{}%
\providecommand \@@endlink[0]{}%
\providecommand \url  [0]{\begingroup\@sanitize@url \@url }%
\providecommand \@url [1]{\endgroup\@href {#1}{\urlprefix }}%
\providecommand \urlprefix  [0]{URL }%
\providecommand \Eprint [0]{\href }%
\providecommand \doibase [0]{https://doi.org/}%
\providecommand \selectlanguage [0]{\@gobble}%
\providecommand \bibinfo  [0]{\@secondoftwo}%
\providecommand \bibfield  [0]{\@secondoftwo}%
\providecommand \translation [1]{[#1]}%
\providecommand \BibitemOpen [0]{}%
\providecommand \bibitemStop [0]{}%
\providecommand \bibitemNoStop [0]{.\EOS\space}%
\providecommand \EOS [0]{\spacefactor3000\relax}%
\providecommand \BibitemShut  [1]{\csname bibitem#1\endcsname}%
\let\auto@bib@innerbib\@empty
\bibitem [{\citenamefont {Rosenau}\ and\ \citenamefont
  {Hyman}(1993)}]{Rosenau:1993zz}%
  \BibitemOpen
  \bibfield  {author} {\bibinfo {author} {\bibfnamefont {P.}~\bibnamefont
  {Rosenau}}\ and\ \bibinfo {author} {\bibfnamefont {J.~M.}\ \bibnamefont
  {Hyman}},\ }\bibfield  {title} {\bibinfo {title} {Compactons: {{Solitons}}
  with finite wavelength},\ }\href {https://doi.org/10.1103/PhysRevLett.70.564}
  {\bibfield  {journal} {\bibinfo  {journal} {Physical Review Letters}\
  }\textbf {\bibinfo {volume} {70}},\ \bibinfo {pages} {564} (\bibinfo {year}
  {1993})}\BibitemShut {NoStop}%
\bibitem [{\citenamefont {Arod{\'z}}(2002)}]{Arodz:2002yt}%
  \BibitemOpen
  \bibfield  {author} {\bibinfo {author} {\bibfnamefont {H.}~\bibnamefont
  {Arod{\'z}}},\ }\bibfield  {title} {\bibinfo {title} {Topological
  compactons},\ }\href@noop {} {\bibfield  {journal} {\bibinfo  {journal} {Acta
  Physica Polonica B}\ }\textbf {\bibinfo {volume} {33}},\ \bibinfo {pages}
  {1241} (\bibinfo {year} {2002})},\ \Eprint
  {https://arxiv.org/abs/nlin/0201001} {arxiv:nlin/0201001} \BibitemShut
  {NoStop}%
\bibitem [{\citenamefont {Arod{\'z}}(2003)}]{Arodz:2003mx}%
  \BibitemOpen
  \bibfield  {author} {\bibinfo {author} {\bibfnamefont {H.}~\bibnamefont
  {Arod{\'z}}},\ }\bibfield  {title} {\bibinfo {title} {Symmetry breaking
  transition and appearance of compactons in a mechanical system},\ }\href
  {https://doi.org/10.48550/arXiv.hep-th/0312036} {\bibfield  {journal}
  {\bibinfo  {journal} {Acta Physica Polonica B}\ }\textbf {\bibinfo {volume}
  {35}},\ \bibinfo {pages} {625} (\bibinfo {year} {2003})},\ \Eprint
  {https://arxiv.org/abs/hep-th/0312036} {arxiv:hep-th/0312036} \BibitemShut
  {NoStop}%
\bibitem [{\citenamefont {Hahne}\ \emph
  {et~al.}(2020{\natexlab{a}})\citenamefont {Hahne}, \citenamefont {Klimas},
  \citenamefont {Streibel},\ and\ \citenamefont {Zakrzewski}}]{Hahne:2019ela}%
  \BibitemOpen
  \bibfield  {author} {\bibinfo {author} {\bibfnamefont {F.~M.}\ \bibnamefont
  {Hahne}}, \bibinfo {author} {\bibfnamefont {P.}~\bibnamefont {Klimas}},
  \bibinfo {author} {\bibfnamefont {J.~S.}\ \bibnamefont {Streibel}},\ and\
  \bibinfo {author} {\bibfnamefont {W.~J.}\ \bibnamefont {Zakrzewski}},\
  }\bibfield  {title} {\bibinfo {title} {Scattering of compact oscillons},\
  }\href {https://doi.org/10.1007/JHEP01(2020)006} {\bibfield  {journal}
  {\bibinfo  {journal} {Journal of High Energy Physics}\ }\textbf {\bibinfo
  {volume} {2020}},\ \bibinfo {pages} {6} (\bibinfo {year}
  {2020}{\natexlab{a}})}\BibitemShut {NoStop}%
\bibitem [{\citenamefont {Hahne}\ and\ \citenamefont
  {Klimas}(2022)}]{Hahne:2022wyl}%
  \BibitemOpen
  \bibfield  {author} {\bibinfo {author} {\bibfnamefont {F.~M.}\ \bibnamefont
  {Hahne}}\ and\ \bibinfo {author} {\bibfnamefont {P.}~\bibnamefont {Klimas}},\
  }\bibfield  {title} {\bibinfo {title} {Compact kink and its interaction with
  compact oscillons},\ }\href {https://doi.org/10.1007/JHEP09(2022)100}
  {\bibfield  {journal} {\bibinfo  {journal} {Journal of High Energy Physics}\
  }\textbf {\bibinfo {volume} {2022}},\ \bibinfo {pages} {100} (\bibinfo {year}
  {2022})}\BibitemShut {NoStop}%
\bibitem [{\citenamefont {Kevrekidis}\ and\ \citenamefont
  {Goodman}(2019)}]{Kevrekidis:2019zon}%
  \BibitemOpen
  \bibfield  {author} {\bibinfo {author} {\bibfnamefont {P.~G.}\ \bibnamefont
  {Kevrekidis}}\ and\ \bibinfo {author} {\bibfnamefont {R.~H.}\ \bibnamefont
  {Goodman}},\ }\href {https://doi.org/10.48550/arXiv.1909.03128} {\bibinfo
  {title} {Four {{Decades}} of {{Kink Interactions}} in {{Nonlinear
  Klein-Gordon Models}}: {{A Crucial Typo}}, {{Recent Developments}} and the
  {{Challenges Ahead}}}} (\bibinfo {year} {2019}),\ \Eprint
  {https://arxiv.org/abs/1909.03128} {arxiv:1909.03128 [nlin]} \BibitemShut
  {NoStop}%
\bibitem [{\citenamefont {Manton}\ \emph
  {et~al.}(2021{\natexlab{a}})\citenamefont {Manton}, \citenamefont {Oleś},
  \citenamefont {Romańczukiewicz},\ and\ \citenamefont
  {Wereszczyński}}]{Manton:2021ipk}%
  \BibitemOpen
  \bibfield  {author} {\bibinfo {author} {\bibfnamefont {N.}~\bibnamefont
  {Manton}}, \bibinfo {author} {\bibfnamefont {K.}~\bibnamefont {Oleś}},
  \bibinfo {author} {\bibfnamefont {T.}~\bibnamefont {Romańczukiewicz}},\ and\
  \bibinfo {author} {\bibfnamefont {A.}~\bibnamefont {Wereszczyński}},\
  }\bibfield  {title} {\bibinfo {title} {Collective coordinate model of
  kink-antikink collisions in $\phi^{4}$ theory},\ }\href
  {https://doi.org/10.1103/PhysRevLett.127.071601} {\bibfield  {journal}
  {\bibinfo  {journal} {Physical Review Letters}\ }\textbf {\bibinfo {volume}
  {127}},\ \bibinfo {pages} {071601} (\bibinfo {year}
  {2021}{\natexlab{a}})}\BibitemShut {NoStop}%
\bibitem [{\citenamefont {Manton}\ \emph
  {et~al.}(2021{\natexlab{b}})\citenamefont {Manton}, \citenamefont {Ole{\'s}},
  \citenamefont {Roma{\'n}czukiewicz},\ and\ \citenamefont
  {Wereszczy{\'n}ski}}]{Manton:2020onl}%
  \BibitemOpen
  \bibfield  {author} {\bibinfo {author} {\bibfnamefont {N.~S.}\ \bibnamefont
  {Manton}}, \bibinfo {author} {\bibfnamefont {K.}~\bibnamefont {Ole{\'s}}},
  \bibinfo {author} {\bibfnamefont {T.}~\bibnamefont {Roma{\'n}czukiewicz}},\
  and\ \bibinfo {author} {\bibfnamefont {A.}~\bibnamefont
  {Wereszczy{\'n}ski}},\ }\bibfield  {title} {\bibinfo {title} {Kink moduli
  spaces: {{Collective}} coordinates reconsidered},\ }\href
  {https://doi.org/10.1103/PhysRevD.103.025024} {\bibfield  {journal} {\bibinfo
   {journal} {Physical Review D}\ }\textbf {\bibinfo {volume} {103}},\ \bibinfo
  {pages} {025024} (\bibinfo {year} {2021}{\natexlab{b}})}\BibitemShut
  {NoStop}%
\bibitem [{\citenamefont {Pereira}\ \emph {et~al.}(2021)\citenamefont
  {Pereira}, \citenamefont {Luchini}, \citenamefont {Tassis},\ and\
  \citenamefont {Constantinidis}}]{Pereira:2020jmi}%
  \BibitemOpen
  \bibfield  {author} {\bibinfo {author} {\bibfnamefont {C.~F.~S.}\
  \bibnamefont {Pereira}}, \bibinfo {author} {\bibfnamefont {G.}~\bibnamefont
  {Luchini}}, \bibinfo {author} {\bibfnamefont {T.}~\bibnamefont {Tassis}},\
  and\ \bibinfo {author} {\bibfnamefont {C.~P.}\ \bibnamefont
  {Constantinidis}},\ }\bibfield  {title} {\bibinfo {title} {Some novel
  considerations about the collective coordinates approximation for the
  scattering of {$\phi^4$} kinks},\ }\href
  {https://doi.org/10.1088/1751-8121/abd815} {\bibfield  {journal} {\bibinfo
  {journal} {Journal of Physics A: Mathematical and Theoretical}\ }\textbf
  {\bibinfo {volume} {54}},\ \bibinfo {pages} {075701} (\bibinfo {year}
  {2021})}\BibitemShut {NoStop}%
\bibitem [{\citenamefont {Pereira}\ \emph {et~al.}(2023)\citenamefont
  {Pereira}, \citenamefont {{dos Santos Costa Filho}},\ and\ \citenamefont
  {Tassis}}]{Pereira:2021gvf}%
  \BibitemOpen
  \bibfield  {author} {\bibinfo {author} {\bibfnamefont {C.~F.~S.}\
  \bibnamefont {Pereira}}, \bibinfo {author} {\bibfnamefont {E.}~\bibnamefont
  {{dos Santos Costa Filho}}},\ and\ \bibinfo {author} {\bibfnamefont
  {T.}~\bibnamefont {Tassis}},\ }\bibfield  {title} {\bibinfo {title}
  {Collective coordinates for the hybrid model},\ }\href
  {https://doi.org/10.1142/S0217751X23500069} {\bibfield  {journal} {\bibinfo
  {journal} {International Journal of Modern Physics A}\ }\textbf {\bibinfo
  {volume} {38}},\ \bibinfo {pages} {2350006} (\bibinfo {year}
  {2023})}\BibitemShut {NoStop}%
\bibitem [{\citenamefont {Adam}\ \emph {et~al.}(2022)\citenamefont {Adam},
  \citenamefont {Manton}, \citenamefont {Oles}, \citenamefont
  {Romanczukiewicz},\ and\ \citenamefont {Wereszczynski}}]{Adam:2021gat}%
  \BibitemOpen
  \bibfield  {author} {\bibinfo {author} {\bibfnamefont {C.}~\bibnamefont
  {Adam}}, \bibinfo {author} {\bibfnamefont {N.~S.}\ \bibnamefont {Manton}},
  \bibinfo {author} {\bibfnamefont {K.}~\bibnamefont {Oles}}, \bibinfo {author}
  {\bibfnamefont {T.}~\bibnamefont {Romanczukiewicz}},\ and\ \bibinfo {author}
  {\bibfnamefont {A.}~\bibnamefont {Wereszczynski}},\ }\bibfield  {title}
  {\bibinfo {title} {Relativistic moduli space for kink collisions},\ }\href
  {https://doi.org/10.1103/PhysRevD.105.065012} {\bibfield  {journal} {\bibinfo
   {journal} {Physical Review D}\ }\textbf {\bibinfo {volume} {105}},\ \bibinfo
  {pages} {065012} (\bibinfo {year} {2022})}\BibitemShut {NoStop}%
\bibitem [{\citenamefont {Adam}\ \emph {et~al.}(2023)\citenamefont {Adam},
  \citenamefont {Ciurla}, \citenamefont {Oles}, \citenamefont
  {Romanczukiewicz},\ and\ \citenamefont {Wereszczynski}}]{Adam:2023qgx}%
  \BibitemOpen
  \bibfield  {author} {\bibinfo {author} {\bibfnamefont {C.}~\bibnamefont
  {Adam}}, \bibinfo {author} {\bibfnamefont {D.}~\bibnamefont {Ciurla}},
  \bibinfo {author} {\bibfnamefont {K.}~\bibnamefont {Oles}}, \bibinfo {author}
  {\bibfnamefont {T.}~\bibnamefont {Romanczukiewicz}},\ and\ \bibinfo {author}
  {\bibfnamefont {A.}~\bibnamefont {Wereszczynski}},\ }\bibfield  {title}
  {\bibinfo {title} {Relativistic moduli space and critical velocity in kink
  collisions},\ }\href {https://doi.org/10.1103/PhysRevE.108.024221} {\bibfield
   {journal} {\bibinfo  {journal} {Physical Review E}\ }\textbf {\bibinfo
  {volume} {108}},\ \bibinfo {pages} {024221} (\bibinfo {year}
  {2023})}\BibitemShut {NoStop}%
\bibitem [{\citenamefont {Blaschke}\ \emph {et~al.}(2023)\citenamefont
  {Blaschke}, \citenamefont {Karp{\'i}{\v s}ek},\ and\ \citenamefont
  {Rafaj}}]{Blaschke:2023mxj}%
  \BibitemOpen
  \bibfield  {author} {\bibinfo {author} {\bibfnamefont {F.}~\bibnamefont
  {Blaschke}}, \bibinfo {author} {\bibfnamefont {O.~N.}\ \bibnamefont
  {Karp{\'i}{\v s}ek}},\ and\ \bibinfo {author} {\bibfnamefont
  {L.}~\bibnamefont {Rafaj}},\ }\bibfield  {title} {\bibinfo {title}
  {Mechanization of a scalar field theory in \$1+1\$ dimensions:
  {{Bogomol}}'nyi-{{Prasad-Sommerfeld}} mechanical kinks and their
  scattering},\ }\href {https://doi.org/10.1103/PhysRevE.108.044203} {\bibfield
   {journal} {\bibinfo  {journal} {Physical Review E}\ }\textbf {\bibinfo
  {volume} {108}},\ \bibinfo {pages} {044203} (\bibinfo {year}
  {2023})}\BibitemShut {NoStop}%
\bibitem [{\citenamefont {Arod{\'z}}\ and\ \citenamefont
  {Klimas}(2005)}]{Arodz:2005}%
  \BibitemOpen
  \bibfield  {author} {\bibinfo {author} {\bibfnamefont {H.}~\bibnamefont
  {Arod{\'z}}}\ and\ \bibinfo {author} {\bibfnamefont {P.}~\bibnamefont
  {Klimas}},\ }\bibfield  {title} {\bibinfo {title} {Chain of impacting
  pendulums as non-analytically perturbed sine-{{Gordon}} system},\ }\bibfield
  {journal} {\bibinfo  {journal} {Acta Physica Polonica B}\ }\textbf {\bibinfo
  {volume} {36}},\ \href {https://doi.org/10.48550/arXiv.cond-mat/0501112}
  {10.48550/arXiv.cond-mat/0501112} (\bibinfo {year} {2005}),\ \Eprint
  {https://arxiv.org/abs/cond-mat/0501112} {arxiv:cond-mat/0501112}
  \BibitemShut {NoStop}%
\bibitem [{\citenamefont {Adam}\ \emph {et~al.}(2017)\citenamefont {Adam},
  \citenamefont {{Sanchez-Guillen}},\ and\ \citenamefont
  {Wereszczynski}}]{Adam:2017pdh}%
  \BibitemOpen
  \bibfield  {author} {\bibinfo {author} {\bibfnamefont {C.}~\bibnamefont
  {Adam}}, \bibinfo {author} {\bibfnamefont {J.}~\bibnamefont
  {{Sanchez-Guillen}}},\ and\ \bibinfo {author} {\bibfnamefont
  {A.}~\bibnamefont {Wereszczynski}},\ }\bibfield  {title} {\bibinfo {title}
  {{{BPS}} submodels of the {{Skyrme}} model},\ }\href
  {https://doi.org/10.1016/j.physletb.2017.04.003} {\bibfield  {journal}
  {\bibinfo  {journal} {Physics Letters B}\ }\textbf {\bibinfo {volume}
  {769}},\ \bibinfo {pages} {362} (\bibinfo {year} {2017})}\BibitemShut
  {NoStop}%
\bibitem [{\citenamefont {Klimas}\ \emph {et~al.}(2018)\citenamefont {Klimas},
  \citenamefont {Streibel}, \citenamefont {Wereszczynski},\ and\ \citenamefont
  {Zakrzewski}}]{Klimas:2018woi}%
  \BibitemOpen
  \bibfield  {author} {\bibinfo {author} {\bibfnamefont {P.}~\bibnamefont
  {Klimas}}, \bibinfo {author} {\bibfnamefont {J.~S.}\ \bibnamefont
  {Streibel}}, \bibinfo {author} {\bibfnamefont {A.}~\bibnamefont
  {Wereszczynski}},\ and\ \bibinfo {author} {\bibfnamefont {W.~J.}\
  \bibnamefont {Zakrzewski}},\ }\bibfield  {title} {\bibinfo {title} {Oscillons
  in a perturbed signum-{{Gordon}} model},\ }\href
  {https://doi.org/10.1007/JHEP04(2018)102} {\bibfield  {journal} {\bibinfo
  {journal} {Journal of High Energy Physics}\ }\textbf {\bibinfo {volume}
  {2018}},\ \bibinfo {pages} {102} (\bibinfo {year} {2018})}\BibitemShut
  {NoStop}%
\bibitem [{\citenamefont {Bogomolny}(1976)}]{Bogomolny:1975de}%
  \BibitemOpen
  \bibfield  {author} {\bibinfo {author} {\bibfnamefont {E.~B.}\ \bibnamefont
  {Bogomolny}},\ }\bibfield  {title} {\bibinfo {title} {Stability of
  {{Classical Solutions}}},\ }\href@noop {} {\bibfield  {journal} {\bibinfo
  {journal} {Sov. J. Nucl. Phys.}\ }\textbf {\bibinfo {volume} {24}},\ \bibinfo
  {pages} {449} (\bibinfo {year} {1976})}\BibitemShut {NoStop}%
\bibitem [{\citenamefont {Prasad}\ and\ \citenamefont
  {Sommerfield}(1975)}]{Prasad:1975kr}%
  \BibitemOpen
  \bibfield  {author} {\bibinfo {author} {\bibfnamefont {M.~K.}\ \bibnamefont
  {Prasad}}\ and\ \bibinfo {author} {\bibfnamefont {C.~M.}\ \bibnamefont
  {Sommerfield}},\ }\bibfield  {title} {\bibinfo {title} {Exact {{Classical
  Solution}} for the 't {{Hooft Monopole}} and the {{Julia-Zee Dyon}}},\ }\href
  {https://doi.org/10.1103/PhysRevLett.35.760} {\bibfield  {journal} {\bibinfo
  {journal} {Physical Review Letters}\ }\textbf {\bibinfo {volume} {35}},\
  \bibinfo {pages} {760} (\bibinfo {year} {1975})}\BibitemShut {NoStop}%
\bibitem [{\citenamefont {Bezanson}\ \emph {et~al.}(2017)\citenamefont
  {Bezanson}, \citenamefont {Edelman}, \citenamefont {Karpinski},\ and\
  \citenamefont {Shah}}]{Bezanson:2017}%
  \BibitemOpen
  \bibfield  {author} {\bibinfo {author} {\bibfnamefont {J.}~\bibnamefont
  {Bezanson}}, \bibinfo {author} {\bibfnamefont {A.}~\bibnamefont {Edelman}},
  \bibinfo {author} {\bibfnamefont {S.}~\bibnamefont {Karpinski}},\ and\
  \bibinfo {author} {\bibfnamefont {V.~B.}\ \bibnamefont {Shah}},\ }\bibfield
  {title} {\bibinfo {title} {Julia: {{A Fresh Approach}} to {{Numerical
  Computing}}},\ }\href {https://doi.org/10.1137/141000671} {\bibfield
  {journal} {\bibinfo  {journal} {SIAM Review}\ }\textbf {\bibinfo {volume}
  {59}},\ \bibinfo {pages} {65} (\bibinfo {year} {2017})}\BibitemShut {NoStop}%
\bibitem [{\citenamefont {Rackauckas}\ and\ \citenamefont
  {Nie}(2017)}]{Rackauckas:2017}%
  \BibitemOpen
  \bibfield  {author} {\bibinfo {author} {\bibfnamefont {C.}~\bibnamefont
  {Rackauckas}}\ and\ \bibinfo {author} {\bibfnamefont {Q.}~\bibnamefont
  {Nie}},\ }\bibfield  {title} {\bibinfo {title} {{{DifferentialEquations}}.jl
  \textendash{} {{A Performant}} and {{Feature-Rich Ecosystem}} for {{Solving
  Differential Equations}} in {{Julia}}},\ }\href
  {https://doi.org/10.5334/jors.151} {\bibfield  {journal} {\bibinfo  {journal}
  {Journal of Open Research Software}\ }\textbf {\bibinfo {volume} {5}},\
  \bibinfo {pages} {15} (\bibinfo {year} {2017})}\BibitemShut {NoStop}%
\bibitem [{\citenamefont {Bazeia}\ \emph {et~al.}(2019)\citenamefont {Bazeia},
  \citenamefont {Mendon{\c c}a}, \citenamefont {Menezes},\ and\ \citenamefont
  {{de Oliveira}}}]{Bazeia:2019tgt}%
  \BibitemOpen
  \bibfield  {author} {\bibinfo {author} {\bibfnamefont {D.}~\bibnamefont
  {Bazeia}}, \bibinfo {author} {\bibfnamefont {T.~S.}\ \bibnamefont {Mendon{\c
  c}a}}, \bibinfo {author} {\bibfnamefont {R.}~\bibnamefont {Menezes}},\ and\
  \bibinfo {author} {\bibfnamefont {H.~P.}\ \bibnamefont {{de Oliveira}}},\
  }\bibfield  {title} {\bibinfo {title} {Scattering of compactlike
  structures},\ }\href {https://doi.org/10.1140/epjc/s10052-019-7519-4}
  {\bibfield  {journal} {\bibinfo  {journal} {The European Physical Journal C}\
  }\textbf {\bibinfo {volume} {79}},\ \bibinfo {pages} {1000} (\bibinfo {year}
  {2019})}\BibitemShut {NoStop}%
\bibitem [{\citenamefont {Arod{\'z}}\ \emph {et~al.}(2006)\citenamefont
  {Arod{\'z}}, \citenamefont {Klimas},\ and\ \citenamefont
  {Tyranowski}}]{Arodz:2005bc}%
  \BibitemOpen
  \bibfield  {author} {\bibinfo {author} {\bibfnamefont {H.}~\bibnamefont
  {Arod{\'z}}}, \bibinfo {author} {\bibfnamefont {P.}~\bibnamefont {Klimas}},\
  and\ \bibinfo {author} {\bibfnamefont {T.}~\bibnamefont {Tyranowski}},\
  }\bibfield  {title} {\bibinfo {title} {Scaling, self-similar solutions and
  shock waves for {{V-shaped}} field potentials},\ }\href
  {https://doi.org/10.1103/PhysRevE.73.046609} {\bibfield  {journal} {\bibinfo
  {journal} {Physical Review E}\ }\textbf {\bibinfo {volume} {73}},\ \bibinfo
  {pages} {046609} (\bibinfo {year} {2006})}\BibitemShut {NoStop}%
\bibitem [{\citenamefont {Hahne}\ \emph
  {et~al.}(2020{\natexlab{b}})\citenamefont {Hahne}, \citenamefont {Klimas},\
  and\ \citenamefont {Streibel}}]{Hahne:2019odw}%
  \BibitemOpen
  \bibfield  {author} {\bibinfo {author} {\bibfnamefont {F.~M.}\ \bibnamefont
  {Hahne}}, \bibinfo {author} {\bibfnamefont {P.}~\bibnamefont {Klimas}},\ and\
  \bibinfo {author} {\bibfnamefont {J.~S.}\ \bibnamefont {Streibel}},\
  }\bibfield  {title} {\bibinfo {title} {Decay of shocklike waves into compact
  oscillons},\ }\href {https://doi.org/10.1103/PhysRevD.101.076013} {\bibfield
  {journal} {\bibinfo  {journal} {Physical Review D}\ }\textbf {\bibinfo
  {volume} {101}},\ \bibinfo {pages} {076013} (\bibinfo {year}
  {2020}{\natexlab{b}})}\BibitemShut {NoStop}%
\bibitem [{\citenamefont {Caputo}\ and\ \citenamefont
  {Flytzanis}(1991)}]{Caputo:1991cv}%
  \BibitemOpen
  \bibfield  {author} {\bibinfo {author} {\bibfnamefont {J.~G.}\ \bibnamefont
  {Caputo}}\ and\ \bibinfo {author} {\bibfnamefont {N.}~\bibnamefont
  {Flytzanis}},\ }\bibfield  {title} {\bibinfo {title} {Kink-antikink
  collisions in sine-gordon and ${\mathrm{\ensuremath{\varphi}}}^{4}$ models:
  Problems in the variational approach},\ }\href
  {https://doi.org/10.1103/PhysRevA.44.6219} {\bibfield  {journal} {\bibinfo
  {journal} {Physical Review A}\ }\textbf {\bibinfo {volume} {44}},\ \bibinfo
  {pages} {6219} (\bibinfo {year} {1991})}\BibitemShut {NoStop}%
\bibitem [{\citenamefont {Takyi}\ and\ \citenamefont
  {Weigel}(2016)}]{Takyi:2016tnc}%
  \BibitemOpen
  \bibfield  {author} {\bibinfo {author} {\bibfnamefont {I.}~\bibnamefont
  {Takyi}}\ and\ \bibinfo {author} {\bibfnamefont {H.}~\bibnamefont {Weigel}},\
  }\bibfield  {title} {\bibinfo {title} {Collective coordinates in
  one-dimensional soliton models revisited},\ }\href
  {https://doi.org/10.1103/PhysRevD.94.085008} {\bibfield  {journal} {\bibinfo
  {journal} {Physical Review D}\ }\textbf {\bibinfo {volume} {94}},\ \bibinfo
  {pages} {085008} (\bibinfo {year} {2016})}\BibitemShut {NoStop}%
\end{thebibliography}%

\end{document}